\documentclass[reprint,
amsmath,amssymb,
bibnotes,
aps,
]{revtex4-2}

\usepackage{graphicx}
\usepackage{dcolumn}

\begin{document}

\title{
Anisotropic Fully-Gapped Superconductivity Possibly Mediated\\
by Charge Fluctuations in a Nondimeric Organic Complex
}

\author{
S. Imajo$^{1,2}$\thanks{imajo@issp.u-tokyo.ac.jp},
H. Akutsu$^2$,
R. Kurihara$^1$,
T. Yajima$^1$,
Y. Kohama$^1$, 
M. Tokunaga$^1$,
K. Kindo$^1$,
and
Y. Nakazawa$^2$
}
\affiliation{
$^1$The Institute for Solid State Physics, the University of Tokyo, Kashiwa, Chiba 277-8581, Japan\\
$^2$Graduate School of Science, Osaka University, Toyonaka, Osaka 560-0043, Japan
}

\begin{abstract}
 We investigate low-temperature electronic properties of the nondimeric organic superconductor $\beta^{\prime\prime}$-(ET)$_4$[(H$_3$O)Ga(C$_2$O$_4$)$_3$]PhNO$_2$.
By examining ultrasonic properties, charge disproportionation (CD) without magnetic field dependence is detected below $T_{\rm CD}$$\sim$8~K just above the superconducting critical temperature $T_{\rm c}$$\sim$6~K.
From quantum oscillations in high fields, we find variation in the Fermi surface and mass enhancement induced by the CD.
Heat capacity studies elucidate that the superconducting gap function is fully gapped in the Fermi surface, but anisotropic with fourfold symmetry.
We point out that the pairing mechanism of the superconductivity is possibly dominated by charge fluctuations.
\end{abstract}

\maketitle

  It is commonly recognized that strong correlations among electrons drive spontaneous self-organization of quantum mechanical degrees of freedom, which results in various kinds of ordered states.
In quarter-filled organic charge-transfer complexes, in which organic molecules are uniformly stacked without dimerization, electron correlations of $\pi$ electrons induce charge instability in itinerant carriers through the intersite Coulomb repulsion $V$.
With the strong correlations, charge density exhibits inhomogeneous distribution in neighboring molecular sites to reduce the Coulomb energy.
Namely it leads to charge ordering (CO) in some geometric patterns as described in the extended Hubbard model\cite{1}.
Indeed, in the electronic phase diagrams of the nondimeric organic compounds, such as $\alpha$-, $\beta^{\prime\prime}$-, $\theta$-type salts\cite{2,2.5,3}, the CO insulator phase is observed in the strong correlation region.
This is in contrast to the dimer-Mott systems like $\kappa$-, $\beta^{\prime}$-, $\rm{\lambda}$-type salts\cite{4,5} where the on-site Coulomb repulsion $U$ gives the Mott physics with magnetic degrees of freedom.
Nevertheless, the electronic phase diagram resembles those of the nondimeric systems in terms of the metal-insulator phase transition driven by the electron correlations.
In such strongly correlated electron systems, superconductivity sometimes appears near the metal-insulator phase boundary, suggesting that the superconductivity originates from the enhanced quantum fluctuations of the electronic degrees of freedom producing the insulating states.
For the dimer-Mott organics\cite{6,7}, the pairing has been intensely studied from viewpoints of antiferromagnetic spin fluctuations and nodal $d$-wave symmetry favored by $U$ although the precise classification in the $d$-wave symmetry is still a controversial issue\cite{6,8,9}.
However, the pairing mechanism in the nondimeric system has not been examined in detail so far.
Some theoretical works for the nondimeric quarter-filled organic systems\cite{10,11,12} suggest that the charge fluctuations developing near the CO phase boundary causes the superconductivity.
Although the experimental results\cite{13,14,15,16,17,18,19,20} imply that the scenario makes sense, there have been no decisive evidences of the relation between the charge instability and the superconductivity.

\begin{figure}
\begin{center}
\includegraphics[width=\hsize]{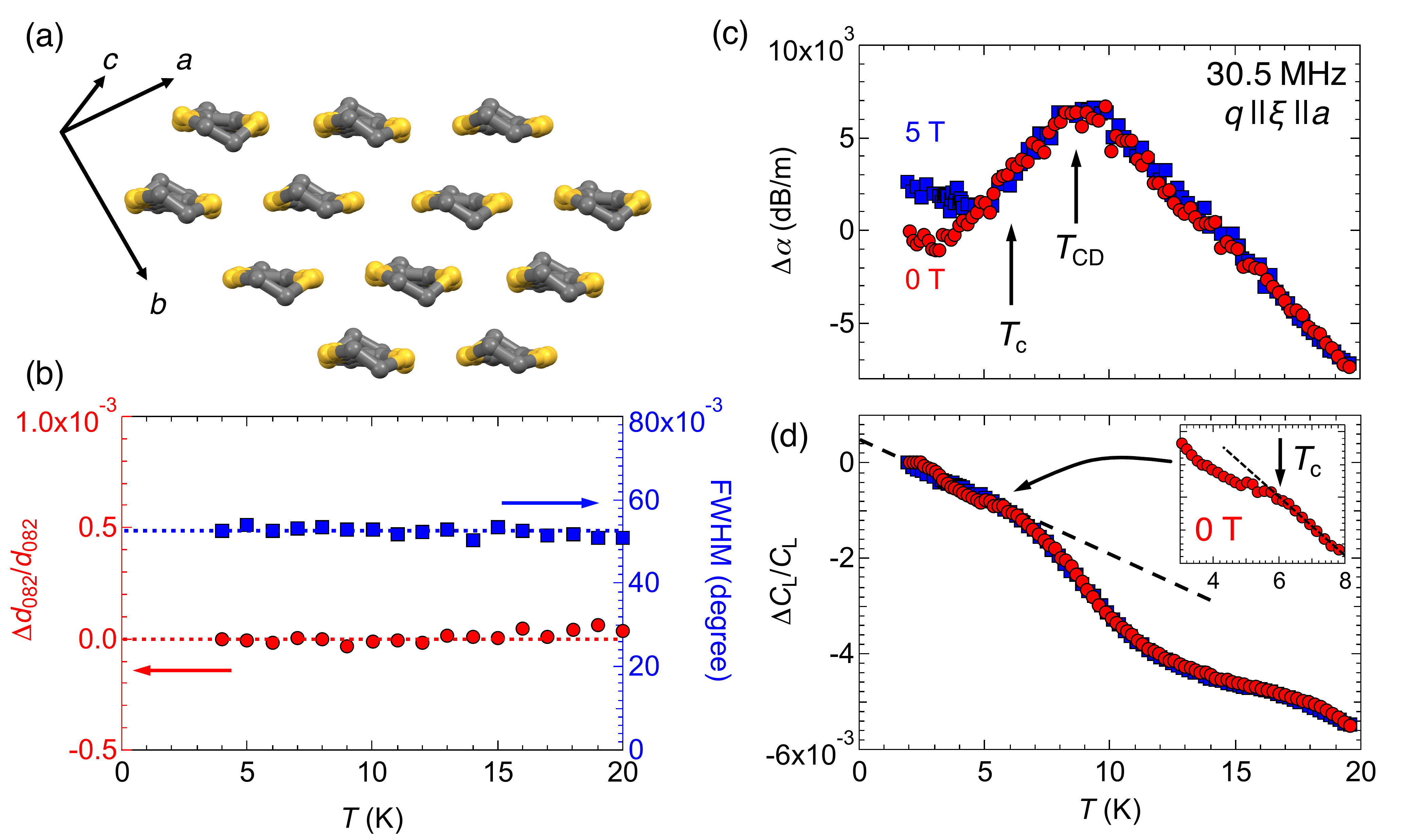}
\end{center}
\caption{
(a) Molecular arrangement of the ET molecules in the conducting plane of $\beta^{\prime\prime}$-(ET)$_4$[(H$_3$O)Ga(C$_2$O$_4$)$_3$]PhNO$_2$.
(b) Temperature dependence of the relative change of the distance between the adjacent (0~8~2) lattice planes $\rm{\Delta}$$d_{082}/d_{082}$ and the FWHM of the (0~8~2) diffraction peak.
(c),(d) Low-temperature ultrasonic properties, (c) the relative change of the ultrasonic attenuation $\rm{\Delta}\alpha$ and (d) the elastic constant ${\rm \Delta} C_{L}/C_{L}$, as a function of temperature.
The arrows indicate the transition temperatures of the CD $T_{\rm CD}$ and superconductivity $T_{\rm c}$.
The red circles and blue boxes denote the data at 0 and 5 T, respectively.
The inset in (d) is the enlarged graph around the superconducting transition temperature $T_{\rm c}$.
}
\label{fig1}
\end{figure}
 A class of charge-transfer salts $\beta^{\prime\prime}$-(ET)$_4$[(H$_3$O)M(C$_2$O$_4$)$_3$]G\cite{20,21,22}, composed of bis-ethylenedithiotetrathiafluvalene (ET) molecule, tris(oxalate)metallate ions [M(C$_2$O$_4$)$_3$$^{3-}$)], and guest molecules G, is known as the example of the nondimeric organics.
The almost uniform arrangement shown in Fig.~\ref{fig1}(a) makes a quarter-filled energy band because of the charge transfer between the counter anion [(H$_3$O)M(C$_2$O$_4$)$_3$]$^{2-}$ and the four organic donors ET$^{0.5+}$.
In our earlier work\cite{20}, we discussed the chemical pressure effect on the electronic states by the chemical substitutions and proposed the electronic phase diagram related to charge instability and electron correlations dominated by $V$.
For one of the salts $\beta^{\prime\prime}$-(ET)$_4$[(H$_3$O)Ga(C$_2$O$_4$)$_3$]PhNO$_2$ which shows a superconductivity transition at $T_{\rm c}$ 6-7~K\cite{21}, the previous studies\cite{13,14,15} report another transition at $\sim$8.5~K just above $T_{\rm c}$, possibly associated with charge disproportionation (CD) due to absence of magnetic ordering and growth of magnetic fluctuations.
The possible CDs have been expected to be the precursor of the CO insulator related to the superconductivity; however, the details of the anomaly are still open.
In this work, we study on $\beta^{\prime\prime}$-(ET)$_4$[(H$_3$O)Ga(C$_2$O$_4$)$_3$]PhNO$_2$ to understand the details of the possible CD and the superconductivity according to the experimental methods in Supplemental Material\cite{Supp},.

 Figure~\ref{fig1}(b) displays the results of the low-temperature x-ray diffraction, the spacing of the lattice plane (0~8~2), $d_{082}$, and the FWHM of the diffraction peak (See Supplemental Material).
The temperature dependence of them indicates the absence of any significant structural change in the measured temperature range 4-20 K.
In Figs.~\ref{fig1}(c) and \ref{fig1}(d), we show the ultrasonic properties, the relative change of the longitudinal ultrasonic attenuation ${\rm \Delta} \alpha$ and the elastic constant ${\rm \Delta} C_{L}/C_{L}$, respectively.
Below 6~K, it shows the magnetic field dependence, considered as the superconducting transition $T_{\rm c}$.
${\rm \Delta} C_{L}/C_{L}$ also exhibits the small anomaly due to the superconducting anomaly at 6~K as shown in the inset of Fig.~\ref{fig1}(d).
Besides, ${\rm \Delta} \alpha$ shows the anomaly around 8~K ($T_{\rm CD}$) as the broad maximum, different from the almost temperature-independent behavior in typical metals.
This should correspond to the possible CD transition reported in the NMR studies\cite{13,14}.
The broad maximum indicates the presence of the development of some fluctuations, which scatters the acoustic phonon and makes the phonon softening observed in ${\rm \Delta} C_{L}/C_{L}$ above $T_{\rm CD}$.
The absence of the magnetic field dependence above $T_{\rm c}$ implies that the anomaly at $T_{\rm CD}$ does not come from the magnetic fluctuations, consistent with the discussion in the earlier NMR studies\cite{13,14}.
Considering the irrelevance to the magnetic and structural instability, the charge fluctuations should make the anomaly since only the charge degree of freedom is left in the present salt.
Thus, these facts demonstrate that the anomaly is predominantly triggered by the charge degrees of freedom.
Indeed, the charge disproportionation can be detected by the ultrasonic properties because charge fluctuations couple well with the strain of lattice induced by ultrasonic waves\cite{Adv.Phys_52}.
This approves the suggested scenario of the realization of the CD at $T_{\rm CD}$, accompanied by the development of the charge fluctuations with approaching $T_{\rm CD}$.

\begin{figure}
\begin{center}
\includegraphics[width=\hsize]{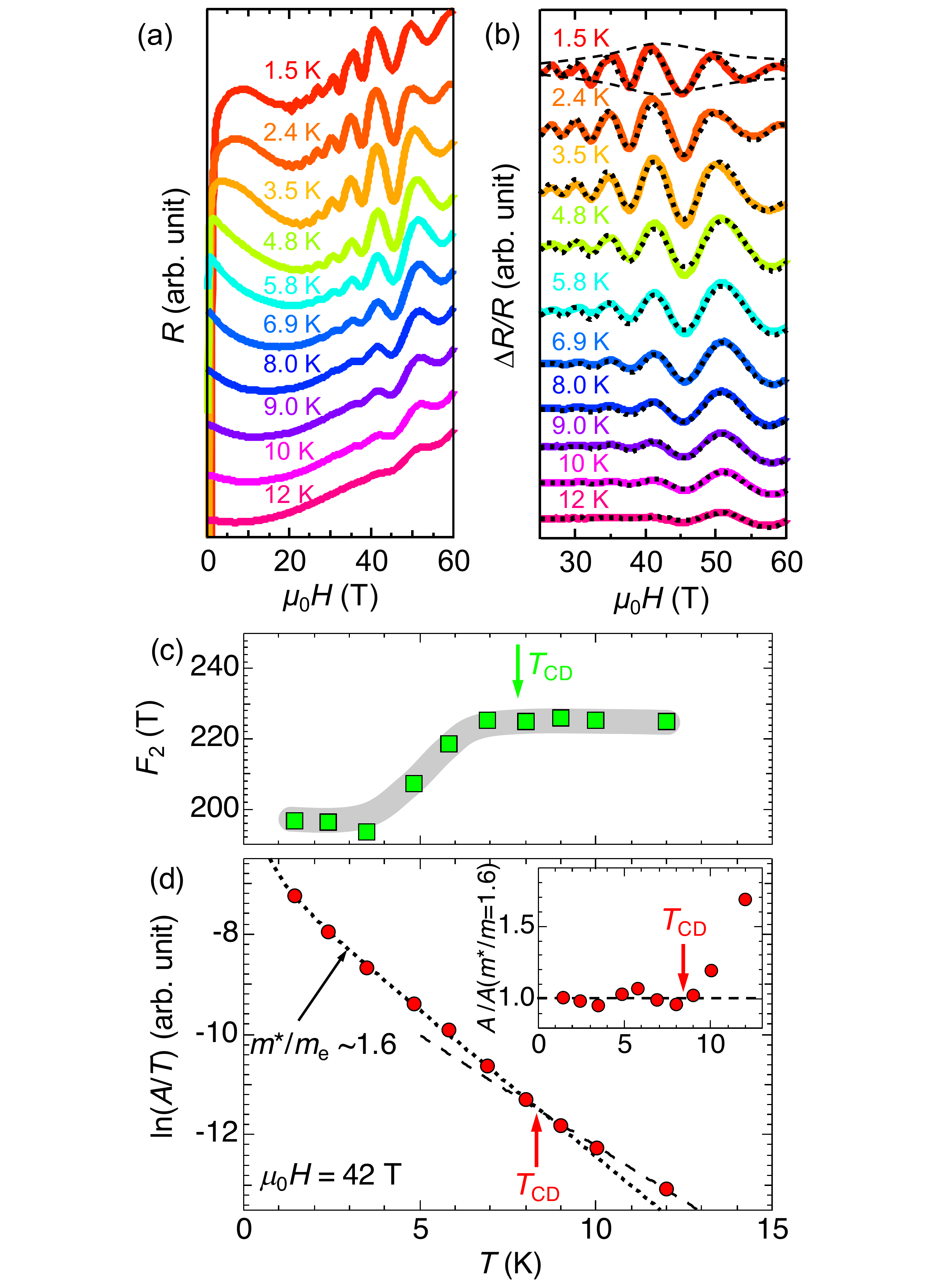}
\end{center}
\caption{
(a) Magnetoresistance at various temperatures.
(b) The oscillatory components above 25~T obtained by subtracting the nonoscillating background from (a).
The dotted curves are fits by using the two-component Lifshitz-Kosevich formula.
The dashed envelopes are guides for the eyes to clarify the amplitude of the oscillations.
(c)Temperature dependences of the frequency of $F_{\rm 2}$.
The gray curve is a guide for the eyes.
(d) A mass plot of the SdH oscillations.
The dotted curve is given by the LK formula with the effective mass $m^{\ast}/m_{\rm e}$=1.6.
The inset displays the ratio of $A$ of the obtained data and the calculation for $m^{\ast}/m_{\rm e}$=1.6, $A/A(m^{\ast}/m_{\rm e}=1.6)$.
}
\label{fig2}
\end{figure}
 Next, we show the results of the electrical transport in the high-field region up to 60~T in Fig.~\ref{fig2}. 
Figure~\ref{fig2}(a) presents the magnetic field dependence of the electrical resistance in magnetic fields applied perpendicularly to the conducting plane.
In the low-field region, the superconductive components can be observed in the curves below 6.9~K, immediately suppressed by applying magnetic fields of a few teslas.
In high fields, the Shubnikov-de Haas (SdH) oscillations are clearly observed from 1.4~K up to 12~K.
In Supplemental Material, we additionally show the results of the high-field magnetic torque measurements.
We confirm that the de Haas-van Alphen (dHvA) oscillations observed in the torque measurements are qualitatively the same with the results of the SdH oscillations.
Figure~\ref{fig2}(b) presents the oscillatory components extracted from the data shown in Fig.~\ref{fig2}(a)\cite{24}.
At higher temperatures, the frequency of the oscillation $F_{\rm 1}$ is determined as $\sim$224~T, which agrees with the previously reported values\cite{20,25,26} and corresponds to the calculated Fermi pockets\cite{20,26}.
Although the quantum oscillations typically become larger in higher fields, the amplitude of the observed oscillations decreases above 40~T at low temperatures as illustrated by the dashed envelopes.
This implies that the oscillation $F_{\rm 1}$ interferes with the new oscillation $F_{\rm 2}$ and the beating of the oscillation occurs at low temperatures.
Indeed, the field dependences can be reproduced by the two-component Lifshitz-Kosevich (LK) formula as described by the dotted curve in Fig.~\ref{fig2}(b).
Here, $F_{\rm 1}$ is fixed at 224~T to reduce the variable parameters because it is difficult to evaluate $F_{\rm 2}$ precisely when $F_{\rm 1}$ is also treated as a variable.
Figure~\ref{fig2}(c) shows the temperature dependence of the frequency of the additional oscillation $F_{\rm 2}$ in the electrical transport.
Such beating generally originates from the warping of the quasi-2D cylindrical Fermi surface due to the interlayer interactions.
However, the present beating manifests only below $T_{\rm CD}$ where the two dimensionality is strongly enhanced\cite{15}.
It means that the mixing of $F_{\rm 2}$ comes from the emergence of the CD.
Since macroscopic phase segregation occurs in the CD state\cite{15}, the overlapping of the quantum oscillations of the CD domains\cite{domain} can give the beating.
As the frequency is proportional to the cross-sectional area of the Fermi pocket, reflecting the number of itinerant carriers, it indicates that the number of carriers of the CD state decreases.
Next, we present the mass plot derived from the amplitude of the original quantum oscillation $F_{\rm 1}$ at $\sim$42~T, $A$, in Fig.~\ref{fig2}(d).
The temperature dependence below $T_{\rm CD}$ gives the effective mass as about 1.6$m_{e}$ as presented by the dotted curve, consistent with the value measured by the low-field ($<$15~T) electrical transport\cite{20}.
On the other hand, above $T_{\rm CD}$ the dwindling tendency with increasing temperature slightly deviates from the fit.
This behavior is obvious in the other plot shown in the inset, the ratio of $A$ of the obtained data and the calculation for $m^{\ast}/m_{e}$=1.6, $A/A(m^{\ast}/m_{e=1.6})$.
The result indicates that the effective electron mass is augmented below $T_{\rm CD}$.
Since the superconductivity is emerged from the anomalous CD metal different from the typical Fermi liquid, the superconductivity is potentially induced by the charge fluctuation.

\begin{figure}
\begin{center}
\includegraphics[width=\hsize,clip]{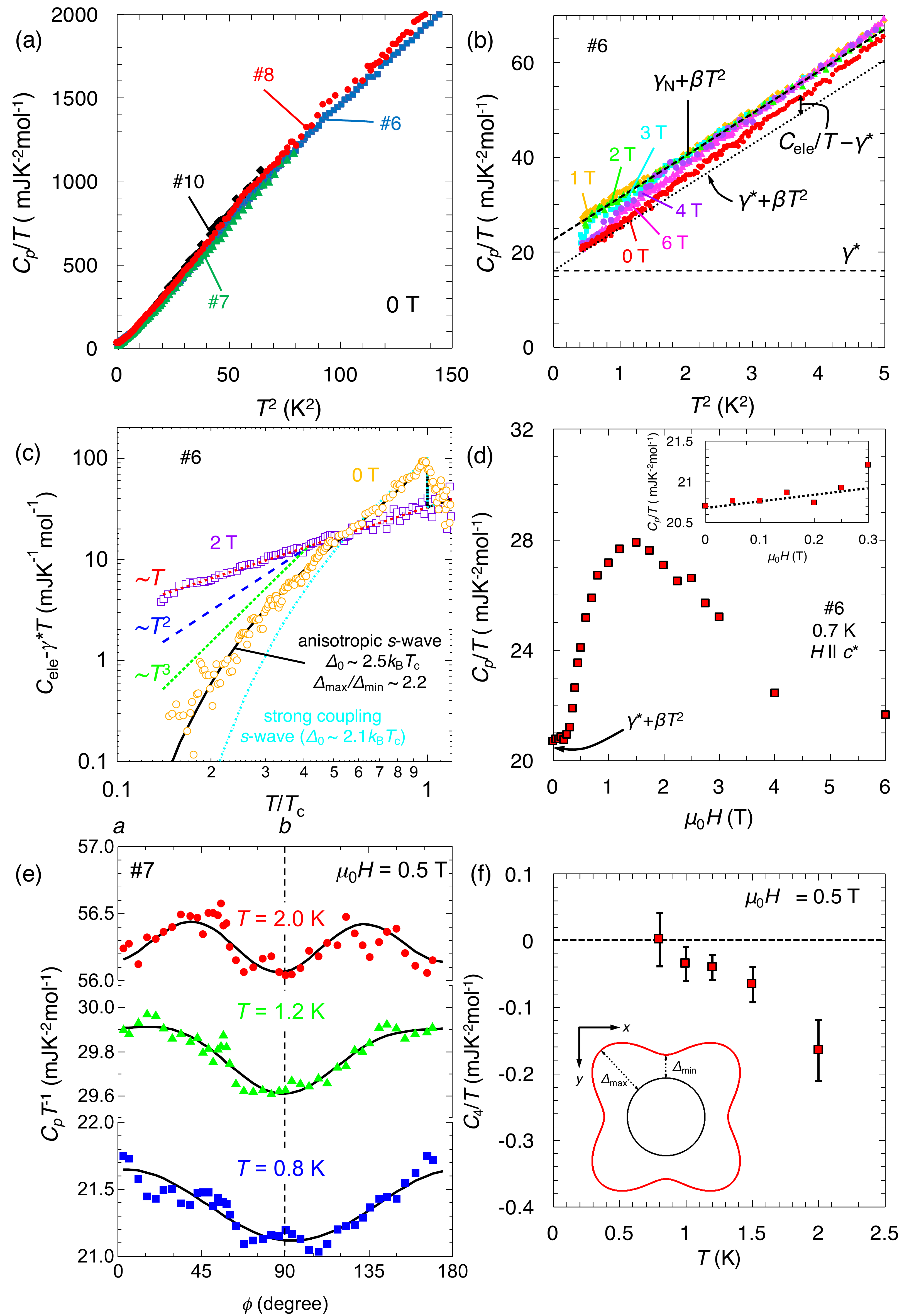}
\end{center}
\caption{
(a) Temperature dependence of zero-field heat capacity plotted as $C_{p}$/$T$ vs. $T^{2}$ for some samples.
(b) Low-temperature heat capacity of sample No. 6 in various magnetic fields.
The lines denote the contribution of the lattice $\beta T^{2}$ and electronic heat capacity $\gamma_{\rm N}$, $\gamma^{\ast}$.
(c) A logarithmic plot of temperature dependent electronic heat capacity at 0 T and 2 T.
The curves represent linear, squared, cubic relations $\sim T^{n}$ ($n$=1-3) and the ${\rm \alpha}$-model calculations for the strong-coupling isotropic $s$-wave superconductivity and the  anisotropic $s$-wave superconductivity.
(d) Magnetic field dependence of the heat capacity at 0.7~K.
The inset is the enlarged plot below 0.3~T.
(e) Variation in $C_{p}$/$T$ for sample No. 7 in an in-plane field 0.5~T rotated from the $a$ axis to the $b$ axis.
The solid curves are the fittings of the equation in the main text.
(f) Temperature dependence of the fourfold component $C_{4}$/$T$.
The negative values mean that the minima of the fourfold term are located at the directions of the crystal axes.
The inset describes a schematic illustration of the angular dependence of the estimated gap function.
}
\label{fig3}
\end{figure}
 In Fig.~\ref{fig3}(a), the zero-field heat capacity for the four different samples is shown in the $C_{p}$/$T$ vs. $T^{2}$ plot\cite{note}.
Only the small sample dependences are found in the present salts and we cannot find the anomaly within the measurement resolution.
Figure~\ref{fig3}(b) is the enlarged plot of the low-temperature region for the sample ${\rm \#}$6 in various magnetic fields applied perpendicular to the conducting plane.
At 0~T, the temperature dependence in this plot gives the finite intercept, corresponding to the residual electronic heat capacity coefficient $\gamma^{\ast}$.
The electronic heat capacity increases with the increase of magnetic field since the normal state recovers from the superconductivity.
The heat capacity of the normal state $C_p$/$T$ = $\gamma_{\rm N}$+$\beta T^{2}$ is derived from the fit in the temperature range between 2 and 4 K$^{2}$, where $\gamma_{\rm N}$ and $\beta$ represent the electronic heat capacity coefficient of the normal state and the lattice heat capacity coefficient, respectively.
The values are obtained as 22.7~mJK$^{-2}$mol$^{-1}$ for $\gamma_{\rm N}$  and 8.9~mJK$^{-4}$mol$^{-1}$ for $\beta$, comparable to those for the typical organic superconductors\cite{27,28}.
In contrast, the ratio $\gamma^{\ast}$/$\gamma_{\rm N}$  reaches 50$\%$-70$\%$, more enormous than 0$\%$-10$\%$ for the typical bulk superconductors\cite{27,28,29}.
This fact means that the electronic state is inhomogeneous even at 0~T.
Since the measured samples are in the clean limit as confirmed by the observation of the quantum oscillation, the phase separation into the superconductivity and normal state is an intrinsic characteristic, which agrees with the results of the EPR study\cite{15} and the beating of the quantum oscillations.
Moreover, the ratio seems to correspond to the intensity ratio of the two NMR peaks in the normal state below $T_{\rm CD}$.
It implies that the superconductivity relates with the phase segregation arising from the CD.
Although it is difficult to clarify the details at present, the mixing of the phase segregation and the second-order-like behavior of the splitting\cite{13,14} might be explained by the recently reported scenario for the bimodal phase coexistence discussed in the inhomogeneous $\alpha$-(BEDT-TTF)$_2$I$_3$\cite{30.5}, which is also the quarter-filled system exhibiting the charge ordering transition.

 For the discussion of the superconductivity, in Fig.~\ref{fig3}(c) we evaluate the electronic heat capacity $C_{\rm ele}$-$\gamma^{\ast}T$ by subtracting the residual electronic heat capacity and lattice heat capacity as a function of reduced temperature $T/T_{\rm c}$.
At 0~T, the low-temperature $C_{\rm ele}$-$\gamma^{\ast}T$ does not vary as $\sim T^{2}$ observed in the line-nodal superconductors including other organic superconductors\cite{27,31,32,33}.
Instead, it seems to obey $\sim T^{3}$ or exponential behavior, indicating that the gap function has point nodes or no nodes in the Fermi surface, respectively.
In order to examine the arisen question regarding whether the gap function has the gap nodes or not, we present the magnetic field dependence of the heat capacity at 0.7~K in Fig.~\ref{fig3}(d) to pursue the quasiparticle excitation in magnetic fields.
At the weak field region below 0.3~T, $C_{p}$/$T$ shows the moderate linear dependence $\sim H$ as shown in the inset.
However, it increases abruptly above 0.3~T and shows a maximum at 1.5~T, which is the critical field $H_{\rm c2}$ in perpendicular fields\cite{26}.
It should be noticed that it decreases in further higher fields.
The field-dependent behavior means that the low-temperature normal state depends on magnetic field in contrast to the case for the typical Fermi liquid although the mechanism is unclear at present.
In the case of the nodal superconductivity, the electronic heat capacity recovers as $\sim H^{0.5}$ for the line-nodal and $\sim H^{0.64}$ for the point-nodal superconductivity in magnetic fields\cite{34} due to the contribution of the quasiparticle excitation at the gap nodes.
On the other hand, fully gapped superconductivity has the quasiparticles excited only inside the vortices, giving $H$-linear dependence in proportion to the number of the vortices.
Therefore, the suppression of the quasiparticle excitation and the linear dependence at low fields strongly indicate the absence of the gap nodes.
Nevertheless, the rapid recovery above 0.3~T similar to the behavior of the nodal superconductivity suggests that the gap symmetry is not a simple $s$-wave.
Here, we assume that the $s$-wave gap has fourfold anisotropy associated with the $d$-wave symmetry.
In fact, the temperature dependence shown in Fig.~\ref{fig3}(c) is reproduced by the solid curve obtained by the ${\rm \alpha}$-model\cite{35} with the fourfold anisotropic $s$-wave gap function $\Delta_{0}$$\sqrt{1+A{\rm cos}(4\phi)}$, where $\Delta_{0}=2.5k_{\rm B}T_{\rm c}$ and $A$=0.64 ($\Delta_{\rm max}$/$\Delta_{\rm min}$=2.2).
The obtained gap amplitude $\Delta_{0}=2.5k_{\rm B}T_{\rm c}$ shows a good agreement with the estimation by the Pauli limiting field\cite{27}.

 Next, we show the in-plane field-angle-resolved heat capacity at the low field 0.5~T ($H$/$H_{\rm c2}<$0.05) in Fig.~\ref{fig3}(e) to confirm the validity of the anisotropic $s$-wave gap function.
The angular dependence is well fitted by the equation $C_{p}$/$T$={$C_{0}$+$C_{2}$cos[2($\phi$-$\phi_2$)]+$C_{4}$cos[4($\phi$-$\phi_4$)]}/$T$ with the fit parameters $\phi_2$ = 2$^{\circ}$$\pm$2$^{\circ}$ and $\phi_4$ = -1$^{\circ}$$\pm$2$^{\circ}$ as presented by the solid curves.
The fourfold component reflects the anisotropic part of the gap function $\sqrt{1+A{\rm cos}(4\phi)}$ while the twofold term comes from the anisotropy of the Fermi velocity influenced by the twofold rotational symmetry of the crystal structure\cite{21}.
Figure~\ref{fig3}(f) shows the temperature dependence of the obtained $C_{4}$/$T$.
With lowering temperature, the fourfold contribution disappears.
As reported in the other anisotropic $s$-wave superconductors\cite{36}, the reduction is exactly the signature of the anisotropic $s$-wave gap because the gap minima do not allow the quasiparticles to be excited by the zero-energy Doppler effect\cite{36,37,38} whereas the nodal superconductivity gives the finite quasiparticle excitations from the gap nodes even in low energy limits.
Hence, we can assert that the gap function is an anisotropic $s$-wave.
Taking account of the finite-energy Doppler shift, the anisotropy is considered as $d_{xy}$-like illustrated in the inset of Fig.~\ref{fig3}(f) since the positions of the gap minima are located at the crystal axis.

 On the basis of the present results, we can claim that the superconductivity occurs in the CD metal and its gap function is an anisotropic $s$-wave.
In the other quarter-filled $\beta^{\prime\prime}$-type compound $\beta^{\prime\prime}$-(ET)$_2$SF$_5$CH$_2$CF$_2$SO$_3$, the full-gap nature\cite{39} and the $d_{xy}$-like anisotropy\cite{40} are reported by heat capacity and transport measurements, respectively.
These features should be consistent with the present results.
Finally, let us discuss the origin of such unconventional superconductivity.
Although there is no specific theoretical study for $\beta^{\prime\prime}$-(ET)$_4$[(H$_3$O)Ga(C$_2$O$_4$)$_3$]PhNO$_2$, some theoretical examinations\cite{10,11,12} of the superconductivity in the proximity to the charge order are reported.
The expected symmetry is $d_{xy}$-\cite{10,11}, $f$-\cite{12}, and $s$-wave\cite{11} for the quarter-filled system, depending on the lattice geometry, $V$, and $U$.
For the case of $U$$\gg$$V$ on the square lattice\cite{11}, $d_{xy}$-symmetry is favored due to the spin fluctuation with the nesting condition.
However, with increasing $V$, the on-site pairing interaction becomes more advantageous instead of the suppression of the pairing around ($\pi$,$\pi$) in the momentum space because of the contribution of the charge fluctuation.
Namely, $s$-wave symmetry can be more stabilized when the charge fluctuation is strong enough.
Thus, the anisotropic $s$-wave symmetry observed in the present work seems to agree with what the superconductivity is mediated by the charge fluctuation.
We still have some unsolved subjects, such as the intrinsic inhomogeneity and the magnetic-field dependent electronic heat capacity of the normal state.
Nevertheless, our results present the decisive discussion of the charge-fluctuation-driven superconductivity.

 In summary, we studied the organic superconductor $\beta^{\prime\prime}$-(ET)$_4$[(H$_3$O)Ga(C$_2$O$_4$)$_3$]PhNO$_2$ from the perspective of the possible charge-fluctuation-mediated superconductivity.
We confirmed that the reported anomaly at $\sim$8~K certainly reflects the development of the charge disproportionation with the growth of the charge fluctuation around $T_{\rm CD}$.
From the analyses of the quantum oscillations observed in high fields, we find that the CD gives rise to the enhancement of the effective mass with the simultaneous decrease of the number of itinerant carriers.
The heat capacity measurements reveal that the unusual CD normal state produces the anisotropic $s$-wave superconductivity, fully gapped with the fourfold anisotropy in the momentum space.
There is no detailed theory about the pairing symmetry of the present superconductivity; however, such unconventional superconductivity possibly originates from the charge fluctuations enhanced in the charge disproportionated metal.

The authors appreciate H. Ishikawa's advice (ISSP, the University of Tokyo) for the low-temperature x-ray diffraction measurement.
The authors thank Y. Nemoto and M. Akatsu (Niigata University) for supplying the LiNbO$_3$ piezoelectric transducers used in this study.

\renewcommand{\thefigure}{S\arabic{figure}}
\clearpage
\begin{center}
\large{\bf{Supplemental Material for\\
Anisotropic Fully-Gapped Superconductivity Possibly Mediated\\
by Charge Fluctuation in a Nondimeric Organic Complex
}}
\end{center}

\section{Experimental methods}
Single crystals of $\beta^{\prime\prime}$-(ET)$_4$[(H$_3$O)Ga(C$_2$O$_4$)$_3$]PhNO$_2$ were electrochemically synthesized\cite{21}.
In-plane and out-of-plane electrical transport measurements were carried out by the typical four-probe method with electric current along the $a$- and $c^{\ast}$-axis, respectively.
While magnetic susceptibility at 1~T was measured by a magnetic property measurement system (Quantum Design), we measured magnetic torque by employing a piezoresistive microcantilever\cite{23}.
Ultrasonic measurements by using the pulse-echo and the phase comparison methods were performed with 31.5~MHz longitudinal ultrasonic waves along the $a$-axis, generated and detected by $36^\circ$ Y-cut LiNbO$_3$ piezoelectric transducers glued on the (1~0~0) surfaces.
Heat capacity including magnetic-field-angle-resolved heat capacity was measured by the reported method using a relaxation calorimeter\cite{7,24}.
These experiments were performed in a $^{3}$He-refrigerator with a 7~T superconducting magnet or a $^4$He-cryostat with a nondestructive 60~T pulse magnet.

\section{Fundamental electrical and magnetic properties}
\begin{figure}[b]
\begin{center}
\includegraphics[width=\hsize,clip]{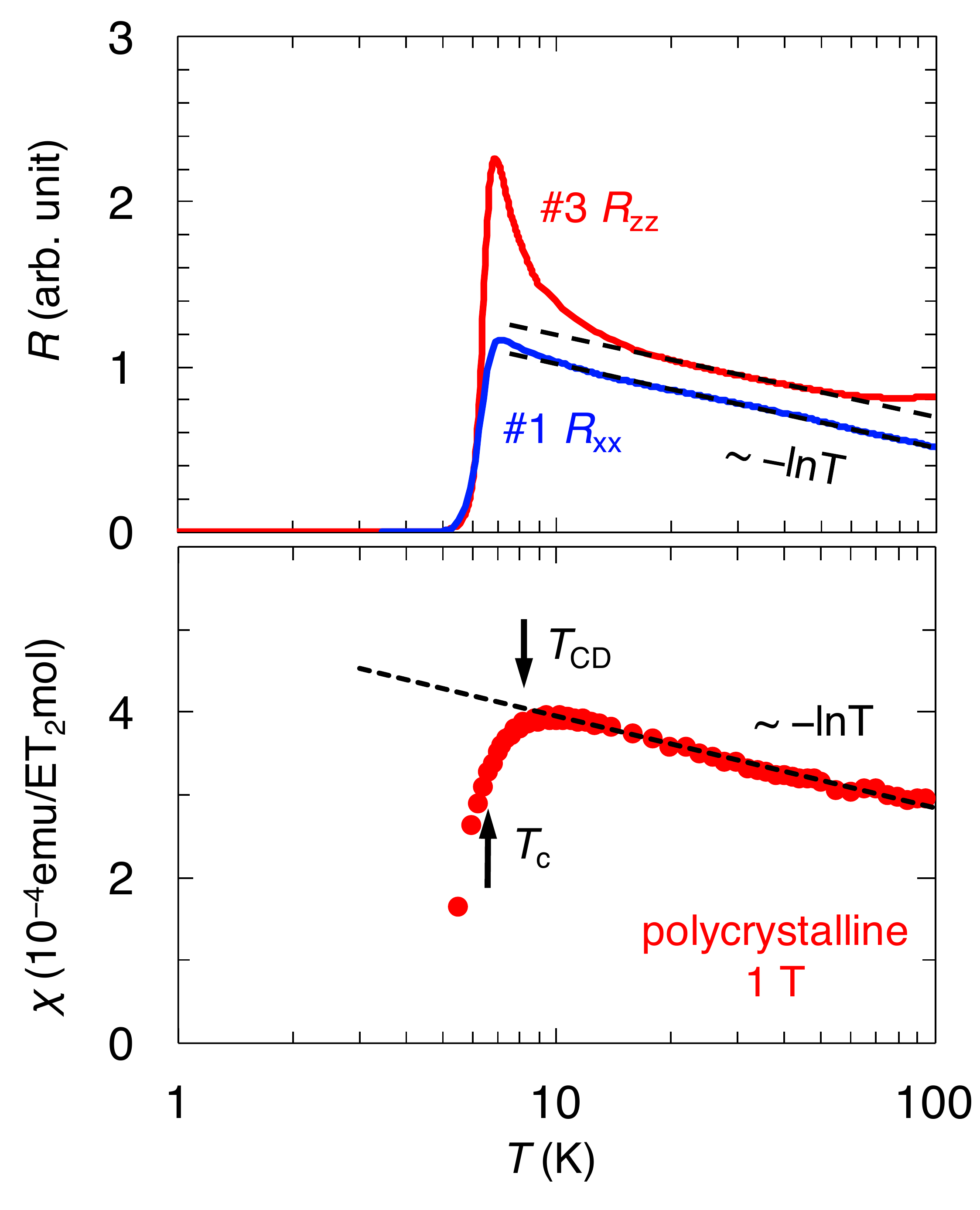}
\end{center}
\caption{
(a),(b) Semi-logarithmic plots of temperature dependences of (a) in-plane and out-of-plane electrical resistance and (b) magnetic susceptibility.
The dashed lines denote $-$ln$T$ dependence.
The arrows in (b) denote the transition temperatures of the charge disproportionation $T_{\rm CD}$ and the superconductivity $T_{\rm c}$.
}
\label{figS1}
\end{figure}
 To confirm the consistency with the results of the previous reports\cite{25,26}, we present the in-plane and out-of-plane electrical resistance as a function of temperature in a semi-logarithmic plot in Fig.~\ref{figS1}(a).
The $-$ln$T$ nonmetallic behavior is commonly observed in a wide temperature range between 10~K and 100~K, whereas only the out-of-plane resistance exhibits the drastic increase with lowering temperature below $\sim$10~K.
The previous report\cite{15} suggests that the discrepancy is explained by the suppression of the interlayer hopping because of the localization of the carriers by the possible CD.
Below 7~K, both the resistance show the steep drop down to zero owing to the emergence of the superconductivity.
The magnetic susceptibility at 1~T as displayed in Fig.~\ref{figS1}(b) also has $-$ln$T$ temperature dependence, which is the same as the electrical transport.
The observed ln$T$ dependence for both is discussed from the viewpoint of the weak localization with scattering centers introduced by the growth of the small CD.\cite{20}
Indeed, the negative magnetoresistance of the normal state in Fig.~2 and Refs.~\cite{25,26,20} can be reproduced by the model based on the weak localization.
The behavior begins to deviate below 8.5~K with a gradual decrease, which should be attributed to the CD.
As reported, the superconductivity appears as the large diamagnetism below 7~K.

\section{Low-temperature x-ray diffraction}
\begin{figure}
\begin{center}
\includegraphics[width=\hsize,clip]{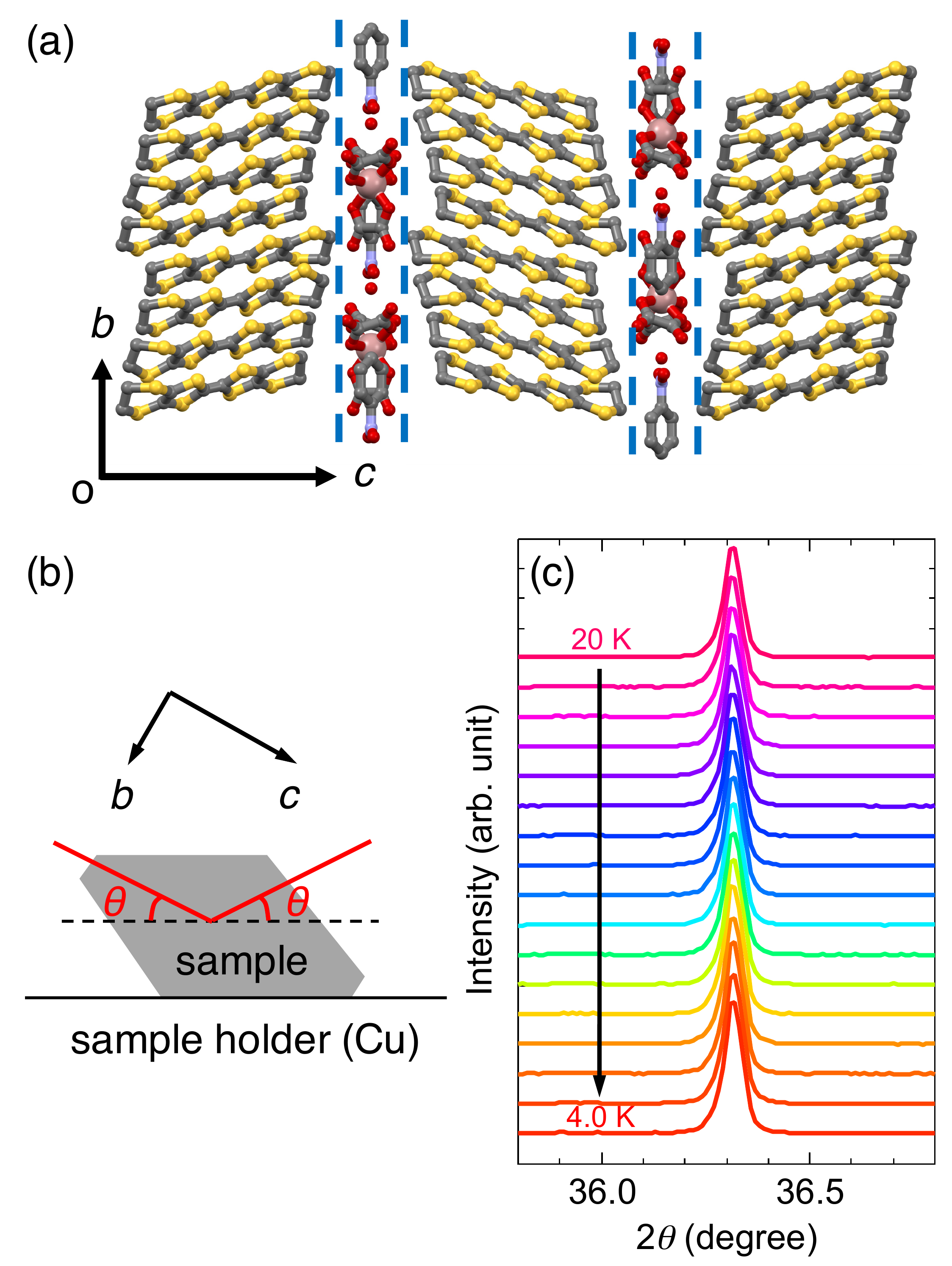}
\end{center}
\caption{
(a) Crystal structure of $\beta^{\prime\prime}$-(ET)$_4$[(H$_3$O)Ga(C$_2$O$_4$)$_3$]PhNO$_2$ viewed along $a$-axis.
The blue dashed lines represent the borders between the ET donor layers and the [(H$_3$O)Ga(C$_2$O$_4$)$_3$]PhNO$_2$ counter-ion counter layers.
(b) A schematic illustration of the powder x-ray diffraction with a single crystal.
The sample is mounted on the Cu sample holder whose surface is parallel to the (0~8~2) lattice planes.
(c) Diffraction peak observed around 2$\theta$ $\sim$36.3 degree in the temperature range from 20 K to 4.0 K.
}
\label{figS2}
\end{figure}
 To acquire the low-temperature structural information, we performed the low-temperature X-ray diffraction.
For typical single-crystal x-ray diffraction, it is quite challenging to cool sample down below 10 K.
In this study, the low-temperature x-ray diffraction of a single crystal was performed in Bragg-Brentano geometry with Cu K$\alpha$1 radiation, monochromated by a Ge(111)-Johansson-type monochromator as illustrated in Fig. ~\ref{figS2}(b), capable of cooling samples down to 4 K.
Although only one lattice plane ($h~k~l$) perpendicular to the sample holder is detectable in this setup, it is enough to discuss whether any structural changes occur at $T_{\rm CD}$ or not.
Figure ~\ref{figS2}(c) is the detected diffraction peak in the temperature range of 4.0 K-20 K.
Considering the orientation and shape of the crystal as well as the lattice parameters at room temperature\cite{21}, the most likely lattice plane is (0~8~2) because it can give a peak at 2$\theta$=36.0 degree at room temperature, close to the peak at 2$\theta$=36.3 degree observed in the present low-temperature X-ray diffraction.
By analyzing the results with the equation $d_{hkl}$=$\lambda$sin$\theta$, where $\lambda$ is the wave length of the used X-ray, and the fits using the Gaussian and Lorentzian functions, we obtain the results displayed in Fig. 1(b) in the main text.
Since this plane has the information of the conducting plane, the present results indicate the absence of the large structural change of the electronic system around $T_{\rm CD}$.

\section{Magnetic torque in pulsed high fields}
\begin{figure}
\begin{center}
\includegraphics[width=\hsize,clip]{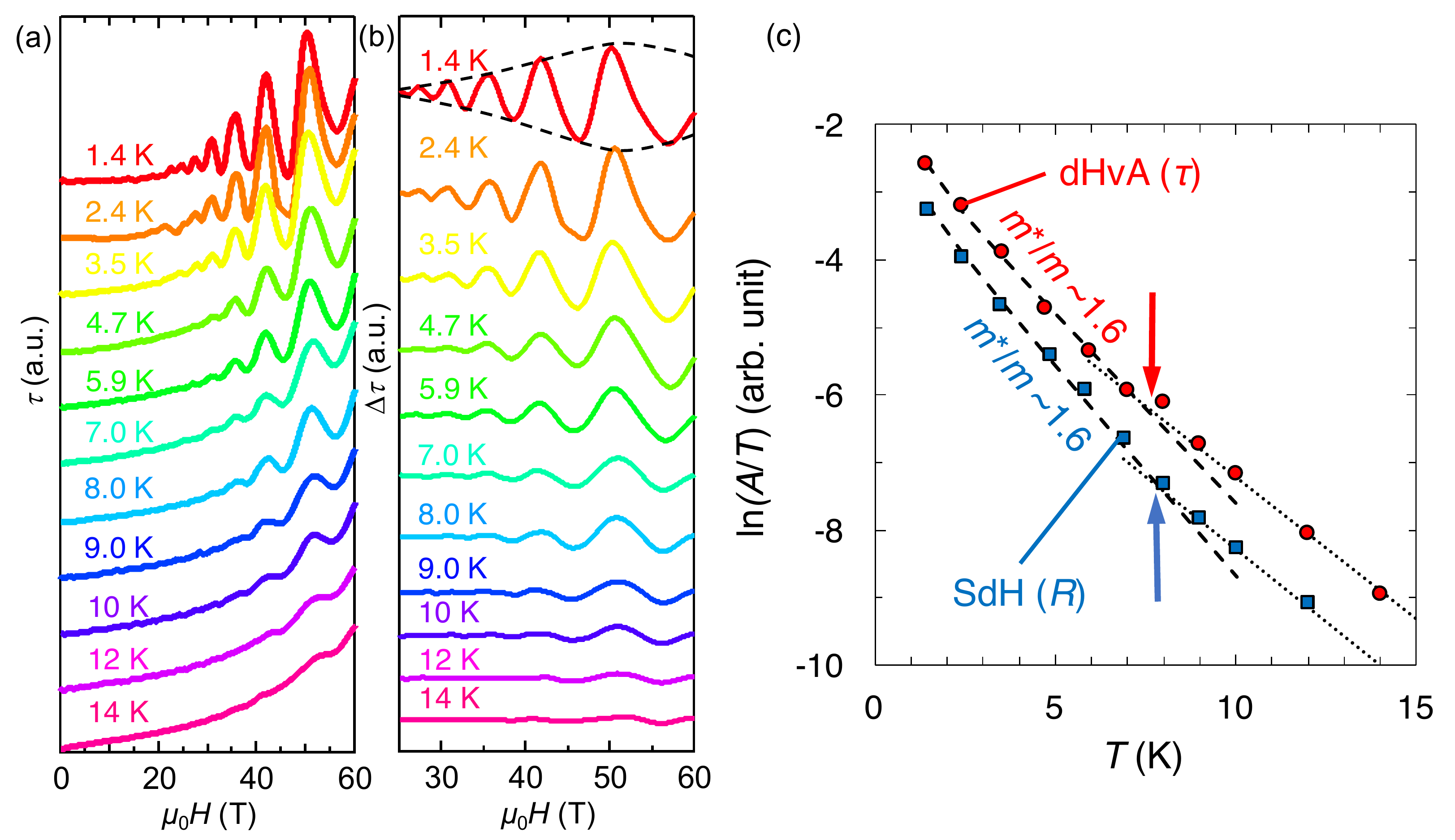}
\end{center}
\caption{
(a) The magnetic field dependence of the magnetic torque up to 60~T at various temperatures.
(b) The oscillatory components above 25~T obtained by subtracting the nonoscillating background from (a)
As in the case of Fig.~2(b) in the main text, the dashed envelopes are guides for the eyes to clarify the amplitude of the oscillations.
(c) The mass plot of the dHvA oscillations with the data of the SdH oscillations shown in Fig. 2(d).
}
\label{figS3}
\end{figure}

 Figure~\ref{figS3}(a) shows the result of the magnetic torque measurement in high fields up to 60~T when magnetic fields are applied almost perpendicular to the two-dimensional conducting plane.
As in the case of the SdH oscillations in the electrical transport, the de Haas-van Alphen (dHvA) oscillations are clearly observed from 1.4~K up to 14~K.
As discussed in the main text, the dHvA oscillations are qualitatively same with the SdH oscillations.
The frequency of the dHvA oscillation below 40~T is about 224~T, which is well consistent with $F_1$ in the SdH oscillation.
The parabolic background should originate from the anisotropy of the magnetic susceptibility of the Pauli paramagnetic component, roughly varies as $\sim H^{2}$.
In Fig.~\ref{figS2}(b), we display the oscillated component of the torque by subtracting the background.
The magnetic field dependence of the oscillation qualitatively agrees with that observed in the SdH in terms of the damping amplitude induced by the mixing of the additional oscillation $F_{2}$. 
As shown in Fig.~\ref{figS2}(c), the results of the SdH and dHvA oscillations are qualitatively same.



\begin{thebibliography}{99}
\bibitem{24} The nonoscillating background is estimated by a polynomial fit to the data points at the nodes of the oscillations.
\bibitem{domain} From the SdH oscillations, we obtain the mean-free path of the measured sample as $\sim$20~nm. Thus, the observation of the SdH oscillations indicates that the domain size of the CD is enough larger than $\sim$20~nm.
\end{thebibliography}
\end{document}